\documentclass[twocolumn,showpacs,preprintnumbers,amsmath,amssymb,prb]{revtex4}

\usepackage{graphicx} 
\usepackage{dcolumn} 
\usepackage{bm}
\usepackage{subfigure} 
\usepackage{float} 
\usepackage{color}

\begin{document}

\title{\bf{ Thermodynamics of the frustrated one-dimensional  spin-$1/2$ Heisenberg ferromagnet in a magnetic field}}

\author{M. H\"{a}rtel}
\author{J. Richter}
\affiliation{Institut f\"{u}r Theoretische Physik, Otto-von-Guericke-Universit\"{a}t Magdeburg, D-39016 Magdeburg, Germany}
\author{D. Ihle}
\affiliation{Institut f\"{u}r Theoretische Physik, Universit\"{a}t Leipzig, D-04109 Leipzig, Germany}

\date{\today}

\begin{abstract}
We calculate the low-temperature thermodynamic quantities 
(magnetization, correlation functions, transverse and longitudinal correlation lengths, spin
susceptibility, and specific heat) of the 
frustrated 
one-dimensional
spin-half $J_1$-$J_2$ Heisenberg ferromagnet, i.e. for $J_2<
0.25|J_1|$,
in an external magnetic field using a 
second-order Green-function
formalism  and full diagonalization of finite systems.
We determine power-law 
relations for the field dependence of the position and the height
of the maximum of the uniform susceptibility.
Considering the specific heat at
at low magnetic fields, two maxima
in its temperature dependence are found.
\end{abstract}

\maketitle
\section{\label{sec:level1}Introduction}
Frustrated low-dimensional spin systems have attracted increasing
attention.\cite{schollwoeck,Lacroix} 
They represent an ideal playground to study the influence of strong (thermal and quantum) 
fluctuations on thermodynamic quantities. 
The one-dimensional (1D) $J_1-J_2$
Heisenberg model with ferromagnetic nearest-neighbor (NN) coupling $J_1$ and
antiferromagnetic next-nearest-neighbor (NNN) coupling $J_2$ has been studied
extensively over the last years, see, e.g.,
Refs.~\onlinecite{rastelli,bursill,tmrg,honecker,krivnov07,krivnov08,hiki,haertel08,zinke09,lauchli09,sirker2010,zinke10,sato}.
It has been pointed out that this 
model is an appropriate starting point to describe 
experimental results for the family of the quasi-1D edge-shared chain
cuprates, such as  
$LiVCuV_4$, $LiCu_2O_2$, $NaCu_2O_2$, $Li_2ZrCuO_4$, and
$Li_2CuO_2$. 
\cite{gibson,matsuda,gippius,ender,drechs1,drechs3,drechs4,park,drechsQneu,malek} 
The corresponding Hamiltonian of this system with an external magnetic field is given by
\begin{equation}\label{hamilop}
  H=J_1\sum_{\langle i,j\rangle} {\bf S}_i{\bf S}_j
+J_2\sum_{[ i,j]} {\bf S}_i{\bf S}_j-h\sum_iS_i^z ,
\end{equation}
where $\langle i,j\rangle$ runs over the NN and $[ i,j ]$ over the NNN
bonds.
For the exchange constants we assume $J_1 <0$ and $J_2 \ge 0$.
In most of these materials quite large values of $J_2$ are realized 
leading to incommensurate spiral in-chain spin correlations at low
temperatures.
Magnetic field effects can be quite
strong,\cite{dmitriev,honecker,hiki,lauchli09,kolezhuk,nishimoto} in particular, if $J_2$
is near the quantum critical point $J_2^c = |J_1|/4$, where the transition between
the ferromagnetic and incommensurate spiral ground state takes place.
On the other hand, some materials considered as 1D $s=1/2$ ferromagnets, 
such as the copper salt
TMCuC,\cite{LW79,DRS82}  the organic magnets p-NPNN\cite{TTN91,TKI92} and
$\beta$-BBDTA$\cdot$GaBr$_4$,\cite{SGM06}  
may have a weak frustrating NNN exchange interaction. 
Moreover, in Refs.~\onlinecite{malek} and \onlinecite{schmitt} $Li_2CuO_2$ and $Li_2ZrCuO_4$
were identified as quasi-1D $s=1/2$ frustrated ferromagnets 
with $J_2\approx 0.2|J_1|$. Although
for $J_2  < J_2^c$ the ground state of the model (\ref{hamilop}) is
ferromagnetic, it has been shown recently\cite{haertel08,haertel10} that
the low-temperature
thermodynamics is strongly influenced by the frustrating $J_2$.
Moreover,  in
Refs.~\onlinecite{magfeld} and \onlinecite{ihle_new} it has been found 
that 
for unfrustrated 1D quantum ferromagnets  at 
weak magnetic fields the specific heat exhibits an additional
low-temperature maximum indicating a separation of two energy scales.
Interestingly, this field-induced low-temperature maximum appears only in 1D
systems and for the small
spin quantum numbers, $s=1/2$ and $s=1$.\cite{magfeld,ihle_new} 
Therefore, it can be considered as a characteristic feature of 1D quantum ferromagnets.

In this paper we consider the interplay between frustration and magnetic field in
1D
$s=1/2$ ferromagnets, i.e. we study the model (\ref{hamilop}) in the
parameter regime $J_2<|J_1|/4$, where the ferromagnetic ground state is realized. 
In difference to previous investigations,\cite{magfeld,ihle_new}  
neither the Bethe ansatz nor the quantum Monte Carlo method can be used.
Hence, we employ (i) a second-order Green-function method (GFM) for infinite
chains and 
(ii) exact diagonalization (ED) of finite chains with $N=16$ and $N=20$ spins
imposing periodic boundary conditions 
to
calculate thermodynamic properties.
It has been shown in Refs.~\onlinecite{haertel08,magfeld} and  \onlinecite{ihle_new}
that both techniques may provide reliable results for the problem under
consideration.    \\

The Green-function technique for 
Heisenberg spin systems used here was initially introduced 
by Kondo and Yamaij in a rotation-invariant formulation.\cite{kondo} It has been 
further developed and successfully applied to low-dimensional quantum systems over the last 
decade.\cite{haertel08,magfeld,ihle_new,rgm_new,antsyg} 
The specific version of the  method appropriate for spin systems in a magnetic
field was developed in Ref.~\onlinecite{ihle_new}. 

The paper is organized as follows.
In Sec.~\ref{kap2} we illustrate the main features of the GFM. In
Sec.~\ref{kap3} we present our results for the magnetization, the susceptibility,
the spin-spin correlation functions, 
the correlation length, and the specific
heat. 
A summary of our results is given in Sec.~\ref{kap4}.

\section{Second-order Green-function theory}\label{kap2}
To calculate the longitudinal and transverse spin correlation functions and
other thermodynamic quantities, 
we use two-time retarded commutator Green functions.\cite{elk} 
First we determine the longitudinal spin correlation functions 
from the Green function $\langle\langle S_q^z;S_{-q}^z\rangle\rangle_\omega=-\chi_q^{zz}\left(\omega\right)$, 
where $\chi_q^{zz}\left(\omega\right)$ is the longitudinal dynamic spin susceptibility. The equation of motion reads
\begin{align}
  \omega^2\langle\langle S_q^z;S_{-q}^z\rangle\rangle_\omega&=M_q^{zz}+\langle\langle -\ddot{S}_q^z;S_{-q}^z\rangle\rangle_\omega,\\
  M_q^{zz}&=-2\sum_{n=1,2}J_nC_n^{-+}\left(1-\cos nq\right),\label{momzz}
\end{align}
where $C_n^{-+}=\langle S_0^-S_n^+\rangle$. To approximate the second derivative $-\ddot{S}_q^z$ we use a decoupling scheme proposed in 
Refs.~\onlinecite{kondo,rgm_new, magfeld, antsyg, ihle_new} which reads 
\begin{align}
  S_i^zS_j^+S_k^-&=\alpha_{jk}^{zz}\langle S_j^+S_k^-\rangle S_i^z,
\end{align}
where for the vertex parameters $\alpha_{jk}^{zz}$ we assume 
$\alpha_{jk}^{zz}\equiv\alpha_{1}^{zz}$ if $j,k$ are NN, and $\alpha_{jk}^{zz}\equiv\alpha_{2}^{zz}$ otherwise. 
We obtain $-\ddot{S}_q^z=(\omega_q^{zz})^2 S_q^z$ and
\begin{equation}\label{greenfunctionzz}
  \chi_q^{zz}=-\langle\langle S_q^z;S_{-q}^z\rangle\rangle_\omega=\frac{M_q^{zz}}{(\omega_q^{zz})^2-\omega^2}
\end{equation}
with
\begin{eqnarray}\label{dispzz}
  (\omega_q^{zz})^2=&&\sum_{n,m(=1,2)}J_nJ_m(1-\cos nq)\\
 &&\qquad\times\left[K_{n,m}^{zz}+4\alpha_n^{zz}C_n^{-+}(1-\cos mq)\right],\nonumber
\end{eqnarray}
where $K_{n,n}^{zz}=1+2(\alpha_2^{zz}C_{2n}^{-+}-3\alpha_n^{zz}C_n^{-+})$, $K_{1,2}^{zz}=
2(\alpha_2^{zz}C_3^{-+}-\alpha_1^{zz}C_1^{-+})$, and $K_{2,1}^{zz}=K_{1,2}^{zz}+4(\alpha_1^{zz}C_1^{-+}-\alpha_2^{zz}C_2^{-+})$. Applying the spectral 
theorem\cite{elk} to the Green function (\ref{greenfunctionzz}), the longitudinal correlation functions of arbitrary range $n$, $C_n^{zz}=\langle S_0^zS_n^z\rangle$, 
can be calculated by\cite{magfeld,ihle_new}
\begin{align}
  C_n^{zz}&=\frac{1}{N}\sum_{q\neq 0}C_q^{zz}e^{iqn}+\langle S^z\rangle^2,\label{Crzz}\\
  C_q^{zz}&=\langle S_{-q}^zS_q^z\rangle=\frac{M_q^{zz}}{2\omega_q^{zz}}\left[1+2n(\omega_q^{zz})\right],\label{Cqzz}
\end{align}
where $n(\omega_q^{zz})=\left(e^{\omega_q^{zz}/T}-1\right)^{-1}$ is the Bose function. For $n=0$, we get the sum rule $C_0^{zz}=1/4$. 
As shown in 
Ref.~\onlinecite{ihle_new}, the isothermal and the uniform static Kubo 
susceptibility $\chi_0^{zz}$ agree at arbitrary fields and temperatures. Using Eqs. (\ref{momzz}), (\ref{greenfunctionzz}), and (\ref{dispzz}) 
we get the relation
\begin{align}\label{vglsus}
  \frac{\partial \langle S^z\rangle}{\partial h}&=-\frac{2}{\Delta^{zz}}\sum_{n=1,2}n^2J_nC_n^{-+}, \\ 
\Delta^{zz}&=\sum_{n,m(=1,2)}n^2J_nJ_mK_{n,m}^{zz}.
\end{align}
Among others, Eq.~(\ref{vglsus}) can be used to determine
the vertex parameters.

To calculate the transverse correlation functions, we use the Green function 
$\langle\langle S_q^+;S_{-q}^-\rangle\rangle_\omega=-\chi_q^{+-}\left(\omega\right)$, 
where $\chi_q^{+-}\left(\omega\right)$ is the transverse dynamic spin susceptibility. Here the equations of motion read
\begin{eqnarray}
  \omega\langle\langle S_q^+;S_{-q}^-\rangle\rangle_\omega&&=2\langle S^z\rangle+\langle\langle i\dot{S}_q^+;S_{-q}^-\rangle\rangle_\omega,\\
  \omega\langle\langle i\dot{S}_q^+;S_{-q}^-\rangle\rangle_\omega&&=M_q^{+-}+\omega\langle\langle -\ddot{S}_q^+;S_{-q}^-\rangle\rangle_\omega.
\end{eqnarray}
The moment $M_q^{+-}=\langle\left[i\dot{S}_q^+,S_{-q}^-\right]\rangle$ is given by
\begin{equation}
  M_q^{+-}=-4\sum_{n=1,2}J_nC_n\left(1-\cos nq\right)+2h\langle S^z\rangle,
\end{equation}
where $C_n=\frac{1}{2}C_n^{-+}+C_n^{zz}$. In $-\ddot{S}_q^+$ we decouple the
products of three operators as
\begin{equation}
  S_i^-S_j^+S_k^+=\alpha_{i,j}^{+-}\langle S_i^-S_j^+\rangle S_k^++\alpha_{i,k}^{+-}\langle S_i^-S_k^+\rangle S_j^+,
\end{equation}
where again for the vertex parameters $\alpha_{i,j}^{+-}$ we assume 
$\alpha_{jk}^{+-}\equiv\alpha_{1}^{+-}$ if $j,k$ are NN, and $\alpha_{jk}^{+-}\equiv\alpha_{2}^{+-}$ otherwise. 
We obtain
\begin{equation}
  -\ddot{S}_q^+=\left[(\omega_q^{+-})^2-h^2\right]S_q^++2hi\dot{S}_q^+
\end{equation}
with
\begin{eqnarray}
    (\omega_q^{+-})^2=&&\sum_{n,m(=1,2)}J_nJ_m(1-\cos nq)\\
 &&\qquad\times\left[K_{n,m}^{+-}+4\alpha_n^{+-}C_n(1-\cos mq)\right],\nonumber
\end{eqnarray}
where $K_{n,n}^{+-}=1+2(\alpha_2^{+-}C_{2n}-3\alpha_n^{+-}C_n)$, $K_{1,2}^{+-}=2(\alpha_2^{+-}C_3-\alpha_1^{+-}C_1)$, 
and $K_{2,1}^{+-}=K_{1,2}^{+-}+4(\alpha_1^{+-}C_1-\alpha_2^{+-}C_2)$. Finally, the resulting Green functions are
\begin{eqnarray}
  \langle\langle S_q^+;S_{-q}^-\rangle\rangle_\omega&&=\sum_{i=1,2}\frac{A_{qi}}{\omega-\omega_{qi}},\label{GFpm}\\
  \langle\langle i\dot{S}_q^+;S_{-q}^-\rangle\rangle_\omega&&=\sum_{i=1,2}\frac{\omega_{qi}A_{qi}}{\omega-\omega_{qi}},\label{GFpm2}
\end{eqnarray}
where
\begin{eqnarray}
  &\omega_{q1,2}=h\pm\omega_q^{+-},\\
  &A_{q1,2}=\langle S^z\rangle\pm \frac{1}{2\omega_q^{+-}}\left(M_q^{+-}-2h\langle S^z\rangle\right).
\end{eqnarray}
As argued in Ref.~\onlinecite{ihle_new}, a divergence in 
the transverse static spin susceptibility $\chi_q^{+-}(\omega=0)$, which
signals a phase transition, could appear if $\omega_{q_c2}=0$, i.e., $\omega_{q_c}^{+-}=h$. Since for nonzero fields, the Heisenberg ferromagnet (\ref{hamilop}) does not describe a phase transition, 
$\chi_q^{+-}(\omega=0)$ has to be finite at all $q$. To ensure this, we require $A_{q_c2}=0$ which results in the regularity condition
\begin{equation}\label{regcon}
  h\langle S^z\rangle=-2\sum_{n=1,2}J_nC_n\left(1-\cos nq_c\right).
\end{equation}
Following Refs.~\onlinecite{ihle_new} and \onlinecite{antsyg}, we assume this condition to be valid also for
larger
fields, where $h>\omega_{q}^{+-}$ for all $q$, to guarantee the continuity of all quantities.\newline
Applying the spectral theorem to the Green functions (\ref{GFpm}) and (\ref{GFpm2}), we obtain the transverse correlation functions $C_n^{-+}=\langle S_0^-S_n^+\rangle=\frac{1}{N}\sum_qC_q^{-+}e^{iqn}$ and $\tilde{C}_n^{-+}=\langle S_0^-i\dot{S}_n^+\rangle=\frac{1}{N}\sum_q\tilde{C}_q^{-+}e^{iqn}$ of arbitrary range $n$ 
with the structure factors
\begin{eqnarray}
  C_q^{-+}=\sum_{i=1,2}A_{qi}n(\omega_{qi}), \tilde{C}_q^{-+}=\sum_{i=1,2}\omega_{qi}A_{qi}n(\omega_{qi}).\label{corr2}
\end{eqnarray}
From the $s=1/2$ operator identity $S_i^-S_i^++S_i^z=\frac{1}{2}$ we find the sum rule
\begin{equation}\label{sumrule1}
  \langle S^z\rangle=\frac{1}{2}-C_0^{-+}.
\end{equation}
Following Ref.~\onlinecite{ihle_new}, a higher-derivative  sum rule can 
be obtained by multiplying $S_i^-$ by $i\dot{S}_i^+$ and expressing the expectation value
by Eq. (\ref{corr2}),
\begin{eqnarray}\label{sumrule2}
  -\frac{1}{N}\sum_q \omega_q^{+-} \sum_{i=1,2}(-1)^i A_{qi} \; n(\omega_{qi})\nonumber\\
=2\sum_{n=1,2}J_n\left(C_n-\frac{1}{2}\langle S^z\rangle\right).
\end{eqnarray}
\begin{figure}
  \includegraphics[height=6cm]{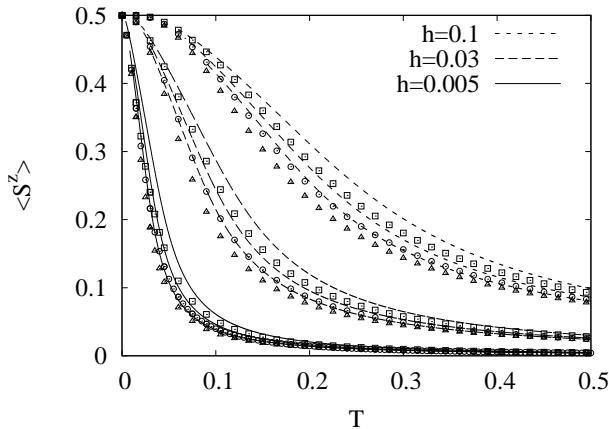}
  \caption{\label{magnetization} 
Magnetization for $J_2=0,0.1,0.15$ (from top to bottom) and $h=0.005,
0.03, 0.1$ versus temperature [GFM - lines; ED ($N=20$) - open symbols].}
\end{figure}
Considering the ground state, at $T=0$, we have the exact results
\begin{equation}
  C_n^{-+}\left(0\right)=0,\qquad C_n^{zz}\left(0\right)=\frac{1}{4}, \qquad \langle S^z\rangle\left(0\right)=\frac{1}{2}.
\end{equation}
Here $C_n^{-+}\left(0\right)$ is independent of $n$, so $A_{q1,2}\left(0\right)$ must be independent of $q$. This requires $K_{n,m}\left(0\right)=0$ which yields 
$\alpha_{1,2}^{+-}(0)=1$. Concerning the zero-temperature values of $\alpha_{1,2}^{zz}$, they can be determined only in the limit $T\to 0$ since 
Eqs. (\ref{Crzz}) and (\ref{Cqzz}) for $C_n^{zz}$ contain $M_q^{zz}$ with $\lim_{T\to 0}M_q^{zz}=0$. To evaluate the thermodynamic properties, 
the correlators $C_n^{zz}$ and $C_n^{-+}$ for $n\neq 0$ are calculated by 
Eqs. (\ref{Crzz}) and (\ref{corr2}), 
whereas $\langle S^z\rangle$ and the 
vertex parameters $\alpha_{1,2}^{+-}$ and $\alpha_{1,2}^{zz}$ are determined 
from the sum rules [Eqs. (\ref{sumrule1}), (\ref{sumrule2}), and
$C_0^{zz}=1/4$], 
the equality (\ref{vglsus}) and the regularity condition (\ref{regcon}).

\section{Results and Discussion}\label{kap3}
In what follows we put $|J_1|=1$. The coupled system of nonlinear algebraic self-consistency equations derived in Sec. \ref{kap2} is solved numerically using 
Broyden's method,\cite{NR2} which yields the solutions 
with a relative error of about $10^{-7}$ on the average.
The momentum sums for $N\to \infty$ are transformed to
integrals which are
evaluated by Gaussian integration.
Tracing the GFM solution to very low temperature we find it to
become less trustworthy for $J_2$ approaching $J_2 = 0.25$. Therefore,
below we will present
GFM results for $J_2 \le 0.2$ only. On the other hand, there is no
restriction with respect to the value of $J_2$ for the ED calculations.  

We 
restrict our discussion to magnetic
fields $h \lesssim 0.1$. For typical values of the exchange parameters of
the order of 10 meV\cite{drechs1,drechs3} that corresponds to 
values of the magnetic field $H \lesssim 10$T mostly used in experiments.

\subsection{Magnetization}
First we consider the magnetization $\langle S^z\rangle$, see Fig. \ref{magnetization}. 
Obviously, at fixed magnetic field the frustration  reduces the magnetization for $T>0$,
whereas
an increase of the magnetic field
at fixed $J_2$ leads to an increase of $\langle S^z\rangle$.
The GFM and ED data  agree well with each other. 

\begin{figure}[ht!]
  \includegraphics[height=11cm]{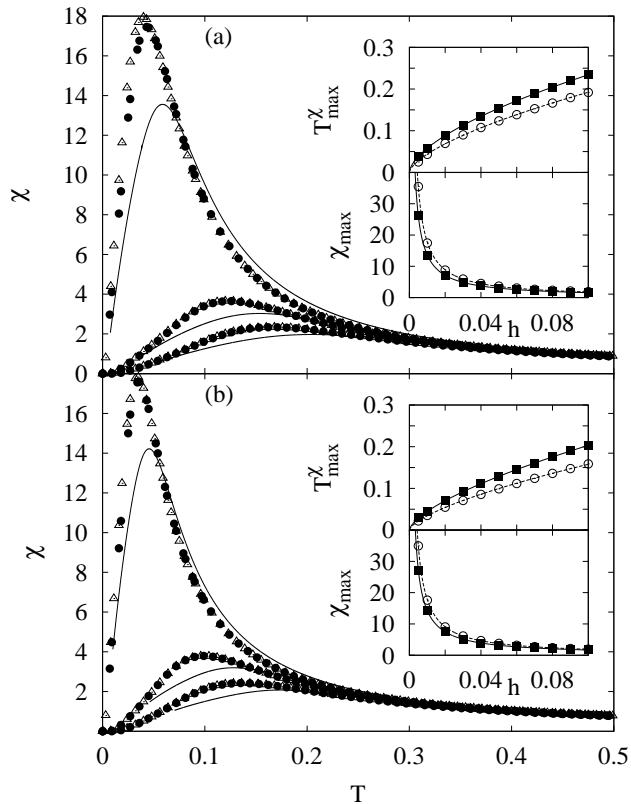}
  \caption{\label{suscept} Magnetic susceptibility $\chi$ versus temperature
  $T$ (GFM -
  lines; ED $N=20$ - filled circles; ED $N=16$ - open triangles) for (a) 
$J_2=0.05$ and (b) $J_2=0.15$ with $h=0.01,0.05,0.08$, from top to bottom. 
The insets show the position and height of the maximum 
obtained by GFM (filled squares) and ED $N=20$ (open circles) as well as the fit by
the power law (lines), see Eq.~(\ref{powerlawchi}).
}
\end{figure}

\subsection{Magnetic susceptibility}
The uniform magnetic susceptibility $\chi=\partial\langle S^z\rangle/\partial h$  at zero field 
diverges at $T=0$ indicating the ferromagnetic phase transition. 
Based on GFM data it was found in Ref.~\onlinecite{haertel08} that 
$\displaystyle \lim_{T \rightarrow 0} \chi T^{2}=(1-4J_2)/24$. (Note that
for $J_2=0$ the rigorous Bethe-ansatz result\cite{yamada} is reproduced.)
In finite magnetic fields  one has $\displaystyle \lim_{T \rightarrow 0}
\chi=0$, and $\chi$ exhibits a maximum. The 
position $T_{max}^\chi$ and the height $\chi_{max}$ 
of this maximum depend on the strength of the magnetic
field $h$ and also on the frustration $J_2$.

The field dependence of the position of the susceptibility maximum
was discussed, for example, in connection with experiments on
La$_{0.91}$Mn$_{0.95}$O$_{3}$.\cite{MRG00} For the unfrustrated ferromagnet
($J_2=0$) 
 it was shown\cite{ihle_new,magfeld,sznaid} that interestingly power laws
for the field dependence of $T_{max}^\chi$ and $\chi_{max}$ are valid, namely
\begin{equation}\label{powerlawchi}
 T_{max}^\chi=ah^{\gamma} \; ,  \qquad  \chi_{max}=bh^{\beta}.
\end{equation}
We mention that our analysis of the susceptibility data
reproduces 
the results reported in  Refs.~\onlinecite{magfeld,ihle_new} 
for $J_2=0$.
Note  that $a, \gamma, b$, and $\beta$ depend on the spin quantum number $s$ and dimensionality
$D$.\cite{magfeld,ihle_new} 

\begin{figure}
  \includegraphics[height=11cm]{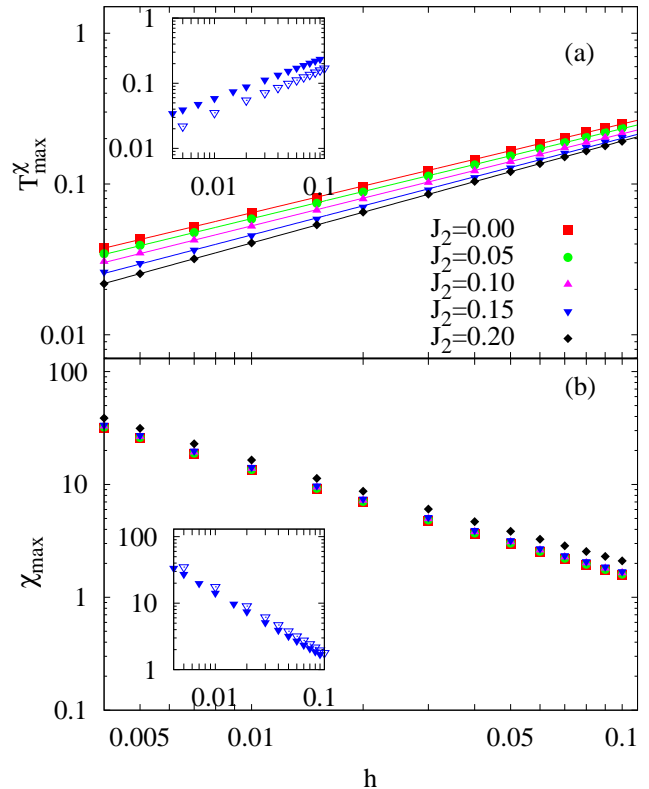}
  \caption{\label{chi_h} Field dependence of (a) the position $T_{max}^\chi$  and
  (b) the height
  $\chi_{max}$  
of the maximum of the magnetic susceptibility 
obtained by GFM (main panels). The lines in the upper panel show the fits by
the power law Eq.~(\ref{powerlawchi}). In the insets the GFM results (filled
triangles) and the corresponding ED data for $N=20$ (open triangles) are
compared for $J_2=0.15$.
}
\end{figure}

\begin{figure}
  \includegraphics[height=11cm]{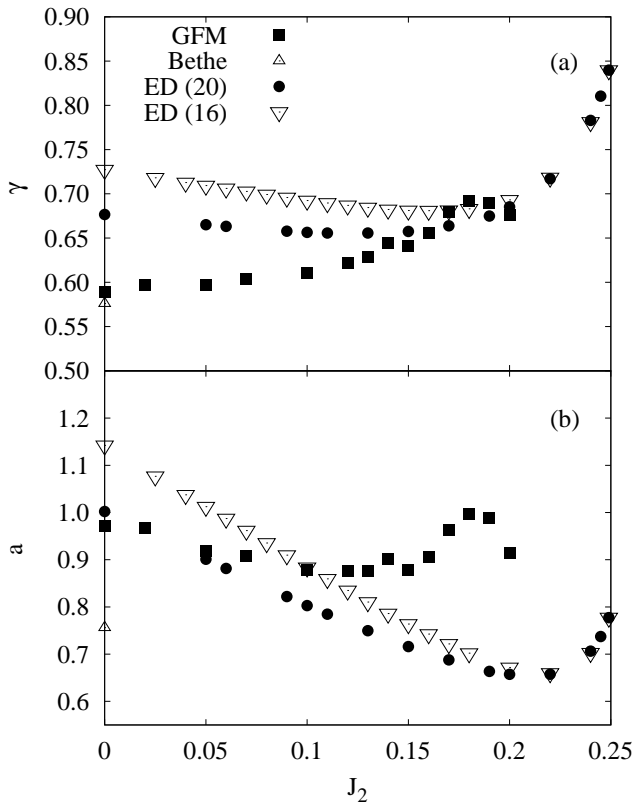}
  \caption{\label{a_alpha} 
 Coefficients (a) $\gamma$ and (b) $a$ 
of the power law
$T_{max}^\chi=ah^{\gamma}$  in dependence on the frustration $J_2$
obtained by GFM and ED (N=16,20). For $J_2=0$ also the Bethe-ansatz 
result\cite{magfeld}  is shown.
}
\end{figure}

\begin{figure}
  \includegraphics[height=11cm]{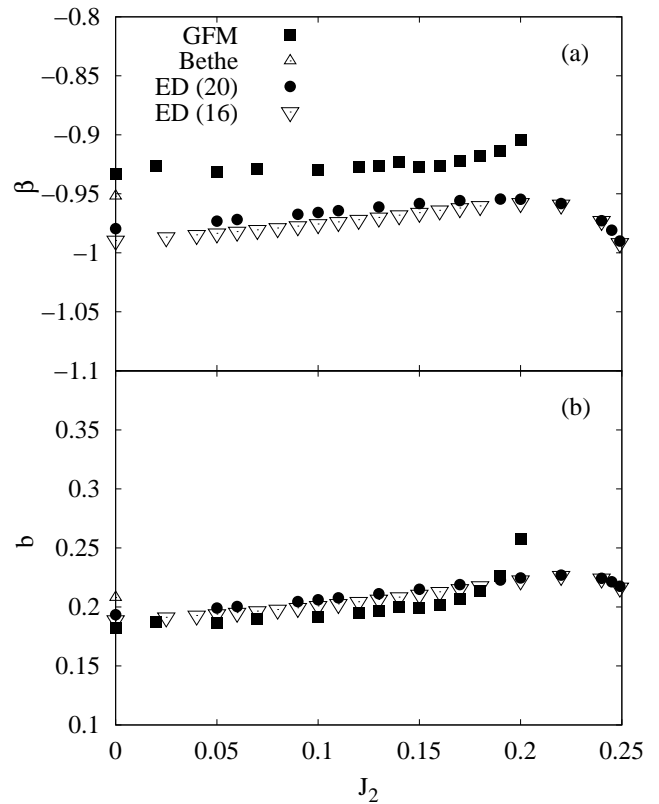}
  \caption{\label{b_beta} Coefficients (a) $\beta$ and (b) $b$ 
of the power law
$\chi_{max}=bh^{\beta}$  in dependence on the frustration $J_2$
obtained by GFM and ED (N=16,20). For $J_2=0$ also the Bethe-ansatz 
result\cite{magfeld}  is shown.
}
\end{figure}

Now we discuss the susceptibility for the frustrated ferromagnet, see Fig.
\ref{suscept}.
The behavior is similar to the unfrustrated case,
i.e. for a fixed $J_2$, 
the position  $T_{max}^\chi$ of the maximum increases and the height
$\chi_{max}$ decreases with increasing field. 
For a fixed $h$ with increasing of frustration $J_2$, the position of the maximum
is slightly shifted to lower temperatures, 
whereas the height is almost fixed for $J_2 < 0.2$.
The ED and GFM data match to each other reasonably well except the
noticeable difference near the maximum.

Next we discuss the validity of the power law (\ref{powerlawchi}) 
for finite frustration $J_2>0$, see insets of Fig. \ref{suscept}. 
We find that the GFM as well as the ED data are well described by the power
law (\ref{powerlawchi}) in the whole range of frustration (i.e. $J_2 \le
0.2$ for the GFM and $J_2 < 0.25$ for the ED), see the log-log plots of  $T_{max}^\chi$ and
$\chi_{max}$ in Fig.~\ref{chi_h}.
The influence of frustration is quite weak. 
To obtain the dependences of the coefficients
on the frustration $J_2$ we fit
the position and the height of the maximum
by the power law using the data points shown in Fig.~\ref{chi_h}.
The variation of the coefficients 
 $a$, $\gamma$, $b$, and $\beta$ with $J_2$ shown in Figs.~\ref{a_alpha} and \ref{b_beta}
  is weak for $J_2< 0.2$ (note the enlarged scale in
Figs.~\ref{a_alpha} and \ref{b_beta}).
The difference between the ED and GFM data may be attributed
to finite-size effects. 
Obviously,
the ED for $N=20$ match better to the GFM data than the ED data for $N=16$
(except for the parameter $a$ at larger $J_2$.)
For comparison, we have also shown the exact Bethe-ansatz
results\cite{magfeld} for $J_2=0$. The GFM data agree well with
the exact data, except for the parameter $a$, where the difference is about
$23\%$. For $J_2 \gtrsim 0.2$, we have only ED data. In this parameter
region the finite size-effects are very small, and there is a
noticeable change of the coefficients $a$, $\gamma$, and $\beta$  with
frustration.

\subsection{Correlation functions}
\begin{figure}
  \includegraphics[height=11cm]{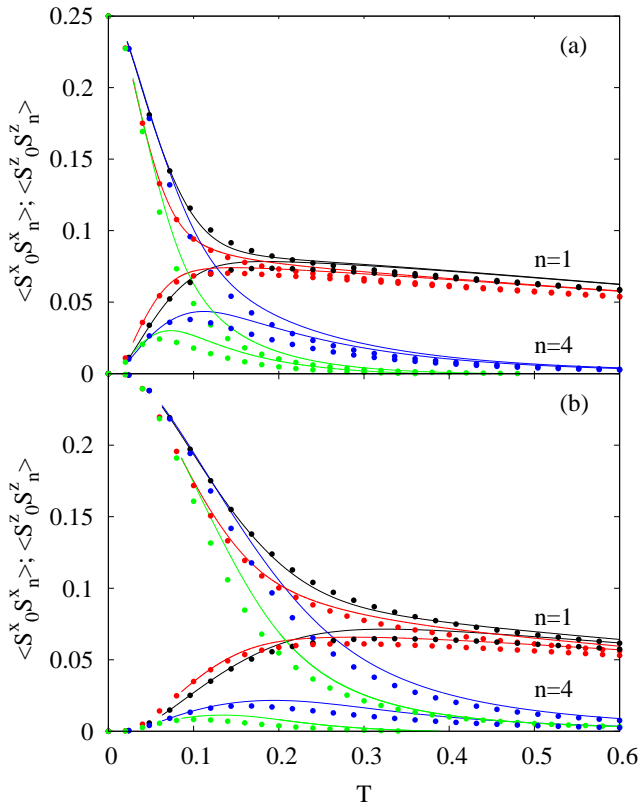}
  \caption{\label{corrfunc} 
Spin-spin 
  correlation functions
$\langle S_0^zS_n^z \rangle$ and $\langle S_0^xS_n^x \rangle$ for $n=1$
and $n=4$
for (a) $h=0.03$ and (b) $h=0.1$ (GFM - lines, ED data for $N=20$ - symbols,
black lines/symbols: $J_2=0$, $n=1$;  red lines/symbols:  $J_2=0.15$,
$n=1$;
blue lines/symbols: $J_2=0$, $n=4$;  green lines/symbols:  $J_2=0.15$,
$n=4$). 
}
\end{figure}

Next we consider the spin-spin correlation functions depicted in Fig. \ref{corrfunc}. 
Obviously, the GFM results for the correlation functions  agree quite well with the ED data.
Due to the magnetic field in $z$-direction we have a different behavior of
the longitudinal correlation functions $\langle S_0^zS_n^z \rangle$ and
the transverse ones $\langle S_0^xS_n^x \rangle=\langle
S_0^yS_n^y \rangle$. 
At $T \gg h$ this difference
disappears. At low temperatures the transverse correlation functions
exhibit a maximum,  where its position  and height depend on the spin-spin
separation
$n$, the frustration $J_2$, and the magnetic field $h$. As expected, frustration
suppresses the correlations, i.e., there is a faster decay of the correlation
functions with increasing $J_2$. 
The strength of the magnetic field at fixed $J_2$ influences the
correlation functions at low temperatures only. Increasing of $h$ yields a
shift of the maximum in $\langle S_0^xS_n^x \rangle$ to higher temperatures and a slower decay
of the correlation functions with $T$.

\subsection{Correlation length}
\begin{figure}
  \includegraphics[height=11cm]{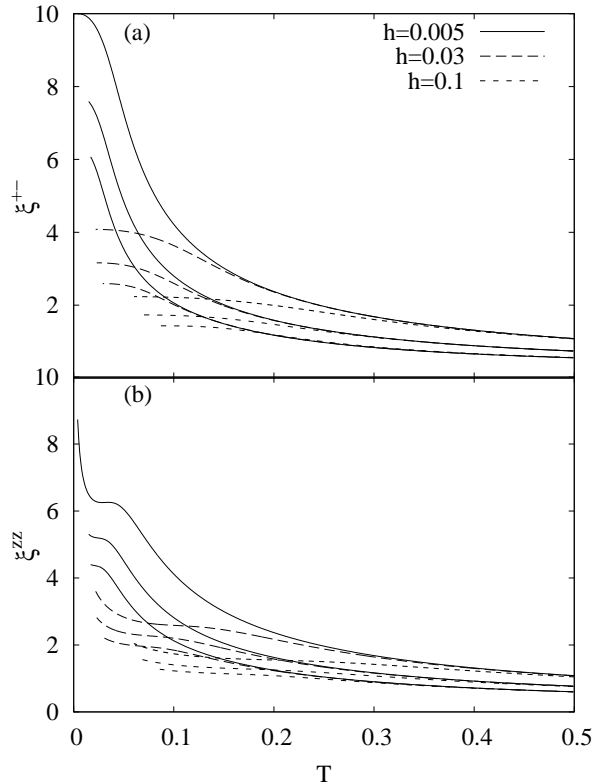}
  \caption{\label{corrlength} (a) Transverse  and (b) longitudinal correlation
  lengths in dependence on the temperature for frustration parameters  $J_2=0,0.1,0.15$ (from top to
  bottom) and  for various values of the magnetic
  field.}
\end{figure}
Within the GFM formalism we are able to calculate the correlation
length.\cite{kondo,haertel08,ihle_new}
Due to the field-induced  anisotropy we have again to distinguish
between the longitudinal   correlation length $\xi^{+-}$
and the transverse correlation length $\xi^{zz}$.
Both quantities can be obtained by 
expanding the static susceptibilities\cite{ihle_new} 
$\chi_q^{\nu\mu}$, $\nu\mu=+-,zz$, around the magnetic wave vector ${\bf q}=0$
which leads to 
$\chi_q^{\nu\mu}=\chi_0^{\nu\mu}/\left\{1+\left(\xi^{\nu\mu}\right)^2q^2\right\}$.
We obtain
\begin{align} \label{chi_pm}
&&  \left(\xi^{+-}\right)^2=\frac{-J_1C_1-4J_2C_2}{\langle S^z\rangle
h}-\frac{\Delta^{+-}}{2h^2} \; ,\\
&& \Delta^{+-}=\sum_{n,m(=1,2)}n^2J_nJ_mK_{n,m}^{+-}
\end{align}
and
\begin{equation} \label{chi_zz}
  \left(\xi^{zz}\right)^2=
\frac{2\left(J_1+4J_2\right)\left(J_1\alpha_1^{zz}C_1^{-+}+4J_2\alpha_2^{zz}C_2^{-+}\right)}{\Delta^{zz}}
,
\end{equation}
where $\Delta^{zz}$ is given by Eq.~(\ref{vglsus}).
As found in  Ref.~\onlinecite{ihle_new}, the transverse and longitudinal correlation
lengths have qualitatively different temperature dependences
at $J_2=0$.
In particular, the longitudinal correlation length
$\xi^{zz}$ reveals an anomaly, namely  a shoulder at $T \sim 0.4$ appearing
at low magnetic fields, cf. also Fig.~\ref{corrlength}(b).

Considering the transverse correlation length $\xi^{+-}$ shown in
Fig.~\ref{corrlength}(a), 
the magnetic field cuts off the 
divergence of the zero-field correlation length at $T=0$ (cf. Ref.~\onlinecite{haertel08}). 
This is related to the absence of the $T=0$ phase transition if $h > 0$.  
From Eq.~(\ref{chi_pm}) we find $\xi^{+-}(T=0)=\sqrt{\frac{0.5-2J_2}{h}}$, 
i.e., at low temperatures
the transverse correlation length increases as $h^{-1/2}$ for $h \to 0$,
and it decreases with increasing frustration according to
$(1-4J_2)^{1/2}$.
At temperatures larger than the magnetic field 
the dependence  on the field strength becomes weak. As already discussed for
the correlation functions, at those temperatures  
both correlation lengths
$\xi^{+-}$ and $\xi^{zz}$ approach each other indicating that the magnetic
field becomes irrelevant. 
On the other hand, the decrease in $\xi^{+-}$ and $\xi^{zz}$  with increasing frustration 
is observed in the whole temperature range shown.
At  low temperatures $T \lesssim h$ 
we find 
$\xi^{zz} < \xi^{+-}$. At very low temperatures this relation might be
reversed. Unfortunately, the GFM does not allow
here a conclusive  statement on $\xi^{zz}$ as $T \to 0$, since 
in Eq.~(\ref{chi_zz}) for $\xi^{zz}$ both the
nominator and denominator go to zero for $T \to 0$ leading to numerical
uncertainties.

\begin{figure}
  \includegraphics[height=11cm]{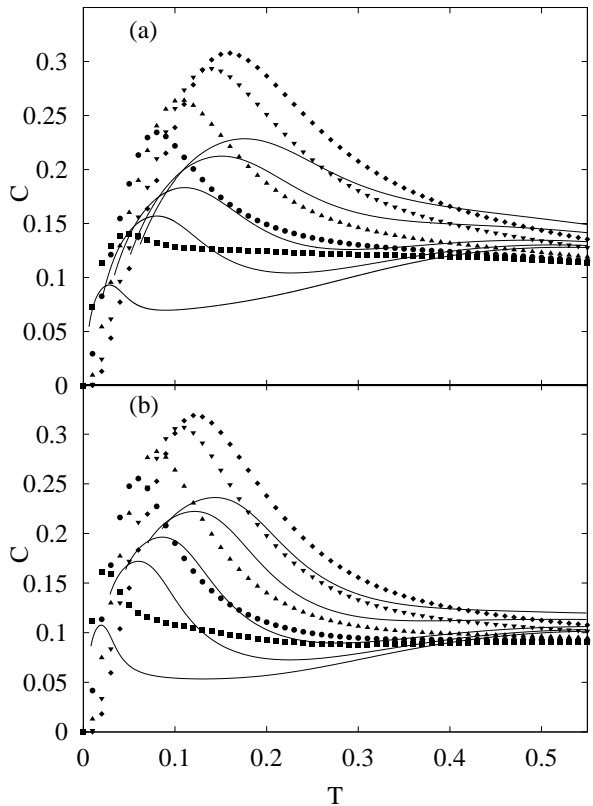}
  \caption{\label{specheat} Specific heat for (a) $J_2=0.05$ and 
(b) $J_2=0.15$ with $h=0.005,0.03,0.05,0.08,0.1$, from bottom to top
(lines - GFM; symbols - ED for $N=20$). }
\end{figure}

\subsection{Specific heat}
Let us first recapitulate some relevant results on the specific heat 
$C(T)=\partial u/\partial T$ with $u=\langle H \rangle/N =
J_1 \left (C_1^{-+}+C_1^{zz} \right )+J_2 \left (C_2^{-+}+C_2^{zz} \right )
-h \langle S^z \rangle$
from previous
investigations.\cite{magfeld,haertel08,tmrg,ihle_new}
For the unfrustrated 1D $s=1/2$ ferromagnet at $h=0$ the specific heat 
shows a broad
maximum at $T \sim 0.35$, see, e.g., Refs.~\onlinecite{magfeld,haertel08}, and
\onlinecite{tmrg}.
In a weak magnetic field $h < h^*$ a double-maximum structure appears. For
the field $h^*$ below which the additional low-temperature maximum appears, a
Bethe-ansatz analysis
gives
$h^*=0.008$, see Ref.~\onlinecite{magfeld}, whereas 
the GFM yields  $h^* \approx 0.07$.\cite{magfeld,ihle_new} 
On the other hand, for the $J_1$-$J_2$ model in zero field also 
a double-maximum structure in $C(T)$ was found. In this case it is induced by
frustration,
and it appears within the GFM for $J_2 > 0.16$.\cite{haertel08} 
Moreover, it was found in Ref.~\onlinecite{haertel08} that, although the
existence of an additional low-temperature maximum can be detected by GFM
and ED, its position and 
height are strongly size dependent, if the maximum is located at very low
temperatures.

Our results for the specific heat of the frustrated ferromagnet in a magnetic
field are shown in Fig.~\ref{specheat}.
Analogously to Refs.~\onlinecite{magfeld} and \onlinecite{ihle_new}, a
double-maximum structure 
is found for small field strengths. 
An increase of the magnetic field shifts the additional low-temperature maximum to 
higher temperatures and increases its height. 
At a sufficiently high field strength $h> h^*$
the typical broad maximum of the pure 
ferromagnet at zero field vanishes, since the low-temperature maximum spreads
into the region of the former broad maximum.    
Although the GFM and the ED data for the specific heat differ
quantitatively, the above mentioned qualitative behavior is found by both
methods., cf. also the discussion in Refs.~\onlinecite{haertel08,magfeld}
and \onlinecite{ihle_new}.
Recalling that for the unfrustrated ferromagnet 
the double-maximum structure disappears for $h > h^* \approx 0.07$ ($h > h^* \approx
0.008$) calculated by GFM (ED), cf. Ref.~\onlinecite{magfeld},
it can be seen in Fig.~\ref{specheat}(b) that $h^*$ is  
shifted to higher values due to frustration. Thus, in  Fig.
\ref{specheat}(b)  
at $J_2=0.15$, the double-maximum structure is still visible in the GFM data at $h=0.08$
and in the ED data at $h=0.03$.  
\begin{figure}
  \includegraphics[height=11cm]{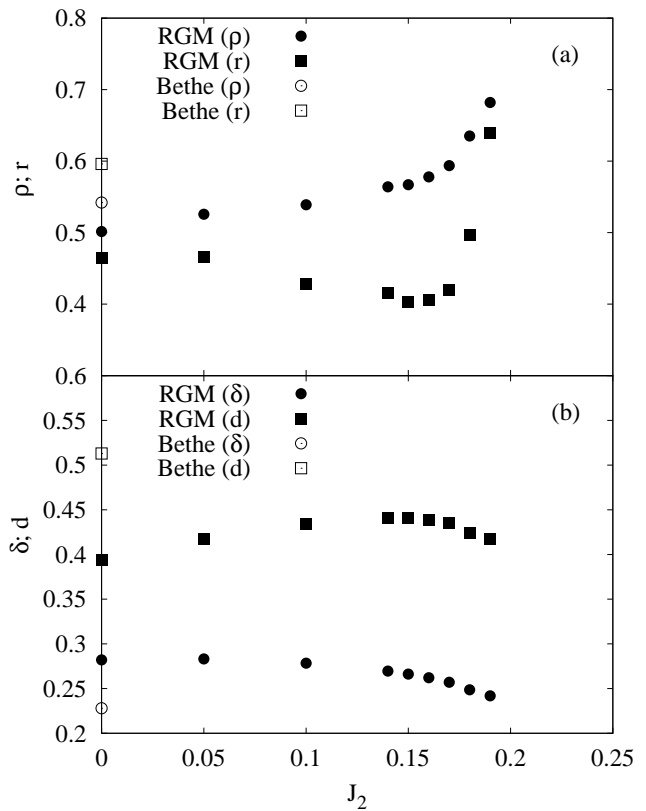}
  \caption{\label{specheatbbeta} (a) Coefficients 
$\rho$  and $r$  of the 
power law $T_{max,1}^C= r h^\rho$
in dependence on the frustration parameter  $J_2$ obtained by GFM. 
For comparison the Bethe-ansatz results are shown for $J_2=0$.
(b) Coefficients 
$\delta$ and $d$  of the 
power law $C_{max,1}=d h^\delta$  in dependence on the frustration parameter  $J_2$ obtained by GFM. 
For comparison the Bethe-ansatz results are shown for $J_2=0$. }
\end{figure}

As pointed out in 
Refs.~\onlinecite{magfeld} and \onlinecite{ihle_new}, 
for $J_2=0$ the position and the height of the additional low-temperature maximum  
follow a power law
\begin{equation}\label{powerlawspecheat}
T_{max,1}^C= r h^\rho \; ,     \qquad  C_{max,1}=d h^\delta.
\end{equation}
To determine  the coefficients $r$,  $\rho$, $d$, and
$\delta$  
in dependence on  the frustration $J_2$ we follow Ref. \onlinecite{ihle_new} and fit 
the position and the height of the low-temperature maximum 
at low fields, $h=0.001-0.01$ in steps of $0.001$, by the power laws of
Eq.~(\ref{powerlawspecheat}). For
these values of the magnetic field, the ED maximum in $C(T)$ quite strongly
depends on the size $N$. Therefore, in Fig.~\ref{specheatbbeta}
we present the GFM results only.
It is obvious that up to $J_2 \sim 0.15$ all coefficients 
depend weakly on $J_2$. Only beyond $J_2\sim 0.15$ the coefficients $r$ and $\rho$
increase noticeably.

Let us finally comment the obvious differences in the 
GFM and ED data for the specific heat, see Fig.~\ref{specheat}. 
The GFM used here and in many previous
papers\cite{haertel08,magfeld,ihle_new,haertel10,kondo,rgm_new,antsyg} 
is a quite universal analytical method which, as a rule,  
cannot yield quantitatively highly accurate data.
On the other hand, the accuracy of the ED data is
also limited at low temperatures  
due to significant finite-size effects, see, e.g.,
the discussion in Ref.~\onlinecite{haertel08}.  
In cases where other accurate methods (for instance  the quantum
Monte Carlo method which is applicable to unfrustrated quantum spin systems,
see e.g.
Ref.~\onlinecite{ihle_new})
are available, it was found that GFM results for the specific
heat at low temperatures show quite large quantitative differences to precise
data, whereas other quantities, such as the susceptibility, are in better
agreement.
However, it has been demonstrated that both the GFM and the ED 
in general yield
qualitatively correct results also for the specific
heat.\cite{haertel08,ihle_new}
In particular, the existence of the 
additional low-temperature maximum in the specific heat  at weak magnetic
fields discussed in this section was confirmed by a Bethe-ansatz
analysis\cite{magfeld} as well as a quantum
Monte Carlo calculation\cite{ihle_new} for the unfrustrated 1D s=1/2
ferromagnet, where both methods are applicable.
Hence, we argue that the existence of the additional low-temperature maximum in the
specific heat is not questioned, however, quanitative statements have to be
taken with caution.

\section{Summary}\label{kap4}
In this paper we use exact diagonalization of finite chains  and a second-order Green's function 
method to study  the influence of a frustrating
next-nearest-neighbor coupling $J_2$ on low-temperature thermodynamic quantities,
such as magnetization, correlation functions, transverse and longitudinal
correlation lengths, spin susceptibility, and specific heat 
of the 
1D $s=1/2$ $J_1$-$J_2$ Heisenberg ferromagnet in a magnetic field $h$. 
We consider $J_2  < 0.25 |J_1|$, i.e., the ground state of the model is
ferromagnetic, but the low-temperature
thermodynamics is strongly influenced by the frustrating $J_2$.
The results of both methods are in good overall agreement. 

The position $T_{max}^\chi$ and the height $\chi_{max}$ 
of  the maximum of the uniform susceptibility appearing in finite magnetic
fields can be described by the power laws  $T_{max}^\chi=ah^{\gamma}$
and $\chi_{max}=bh^{\beta}$, 
where the coefficients $a$, $\gamma$, $b$, and $\beta$
weakly depend on $J_2$. 
The double-maximum structure of the specific heat $C(T)$ found for $J_2=0$ at low 
fields\cite{magfeld,ihle_new} occurs also for $J_2\neq 0$. 
The field strength at which this structure vanishes  
is slightly  increasing with frustration. 
The position and the height of the additional low-field low-temperature 
maximum in $C(T)$ in dependence on the field strength  can be also described 
by power laws.  \\

{\bf Acknowledgment}\\
This work was supported by the DFG (project RI615/16-1).
The full diagonalization of finite spin systems was done using J. Schulenburg's
{\it spinpack.}


\begin{thebibliography}{99}
\bibitem{schollwoeck} Quantum Magnetism, eds. U. Schollw\"{o}ck, J. Richter, D. J. J. Farnell, and R. F. Bishop, 
Lecture Notes in Physics {\bf{645}} (Springer-Verlag, Berlin, 2004).
\bibitem{Lacroix} 
Introduction to Frustrated
Magnetism, eds.
C. Lacroix, P. Mendels and F. Mila, 
Springer Series in
      Solid-State Sciences {\bf 164} (Springer-Verlag, Berlin, 2011).


\bibitem{rastelli}A.\ Pimpinelli, E.\ Rastelli, and A.\ Tassi, { J. Phys.:
Condens. Matter} {\bf 1}, 7941
(1989).


\bibitem{bursill}
R.~Bursill, G.A.~Gehring, D.J.J.~Farnell, J.B.~Parkinson,  T.~Xiang, and
C.~Zeng,
{ J. Phys.: Condens. Matter} {\bf 7}, 8605 (1995).

\bibitem{tmrg} H. T. Lu, Y. J. Wang, Shaojin Qin, and T. Xiang,
Phys. Rev. B \textbf{74}, 134425 (2006).


\bibitem{honecker}F.\ Heidrich-Meisner, A.\ Honecker, and T.\ Vekua, 
Phys.\ Rev.\ B {\bf 74} 020403(R) (2006).

\bibitem{krivnov07} D.V.~Dmitriev, V.Ya.~Krivnov, and J.~Richter,
       Phys. Rev. B {\bf 75}, 014424 (2007).
\bibitem{krivnov08}
D.V. Dmitriev and V.Ya. Krivnov,
Phys. Rev. B {\bf 77}, 024401 (2008).

\bibitem {hiki} T. Hikihara, T. Momoi , A. Furusaki, and  H. Kawamura  Phys.
Rev. B {\bf 78}, 144404 (2008).
 
\bibitem {haertel08} M. H\"{a}rtel, J. Richter, D. Ihle, and  S.-L.
Drechsler,
         Phys. Rev. B {\bf 78}, 174412 (2008);
J. Richter, M. H\"{a}rtel, D. Ihle, and  S.-L. Drechsler,
       J. Phys.: Conf. Ser. {\bf 145}, 012064 (2009). 


\bibitem{zinke09}  R. Zinke,  S.-L. Drechsler, and J. Richter,
       Phys. Rev. B {\bf 79}, 094425 (2009).


\bibitem{lauchli09} J. Sudan, A. Luscher, and A. L\"auchli,
Phys. Rev. B {\bf 80}, 140402(R) (2009). 


\bibitem{sirker2010}  J. Sirker,
       Phys. Rev. B {\bf 81}, 014419  (2010).

\bibitem{zinke10} 
 R. Zinke, J. Richter, and S.-L. Drechsler,
     J. Phys.: Condens. Matter {\bf 22}, 446002 (2010).

\bibitem{sato} M. Sato, T. Momoi, and A. Furusaki
 Phys. Rev. B {\bf 79}, 060406(R) (2009);
M. Sato, T. Hikihara, and T. Momoi,
Phys. Rev. B {\bf 83}, 064405 (2011).


\bibitem{gibson} B. J. Gibson, R. K. Kremer, A. V. Prokofiev, W. Assmus, and G. J. McIntyre, Physica B: Condensed Matter, {\bf 350} E253 (2004).
\bibitem{matsuda} T.  Matsuda, A. Zheludev, A. Bush, M. Markinka, and A. Vasiliev   Phys. Rev. Lett. {\bf 92}, 177201 (2004).
\bibitem{gippius}  A. A. Gippius, E. N. Morozova1 A. S. Moskvin, A. V. Zalessky, A. A. Bush, M. Baenitz, H. Rosner, and S.-L. Drechsler, Phys. Rev. B {\bf 70}, 020406(R) (2004).
\bibitem{ender} M.~Enderle, C.~Mukherjee, B.~Fak, R.K.~Kremer, J.-M.~Broto, H.~Rosner, S.-L.~Drechsler, J.~Richter, J.~Malek, A.~Prokofiev,W.~Assmus, S.~Pujol, J.-L.~Raggazoni, H.~Rakato, M.~Rheinst\"adter, and H.M.~Ronnow, Europhys. Lett. {\bf 70}, 237 (2005).
\bibitem{drechs1} T.~Masuda, A.~Zheludev, A.~Bush, M.~Markina, and A.~Vasiliev, { Phys. Rev. Lett.} {\bf 92}, 177201 (2004); S.-L.~Drechsler, J.~M\'alek, J.~Richter, A.S.~Moskvin, A.A.~Gippius, and H.~Rosner, { Phys. Rev. Lett.} {\bf 94}, 039705 (2005).
\bibitem{drechs3} S.-L. Drechsler, J. Richter, A.A.\ Gippius, A.\ Vasiliev, A.S.\ Moskvin, J.\ M\'alek, Y.\ Prots, W.\ Schnelle, and H. Rosner, Europhys. Lett. {\bf 73}, 83 (2006).
\bibitem{drechs4}   S.-L.~Drechsler, J.~Richter, R.~Kuzian, J.~M\'alek, N.~Tristan, B.~B\"uchner, A.S.~Moskvin, A.A.\ Gippius, A.\ Vasiliev, O.~Volkova, A.~Prokofiev, H.~Rakato, J.-M.~Broto, W.~Schnelle, M.~Schmitt, A.~Ormeci, C.~Loison, and H.~Rosner,  { J. Magn. Magn. Mater.} {\bf 316}, 306 (2007).
\bibitem{park} S.~Park, Y.J.~Choi, C.L.~Zhang, and S.-W.~Cheong, { Phys. Rev. Lett.} {\bf 98}, 057601 (2007).
\bibitem{drechsQneu}S.-L. Drechsler, O. Volkova, A.N. Vasiliev, N. Tristan, J. Richter, M. Schmitt, H. Rosner, J. M\'alek, R. Klingeler, A.A. Zvyagin, and B. B\"uchner, { Phys. Rev. Lett.} {\bf 98}, 077202 (2007).
\bibitem{malek}J. M\'alek, S.-L. Drechsler, U.~Nitzsche, H. Rosner, and H.~Eschrig, 
Phys. Rev. B {\bf 78}, 060508 (2008).
\bibitem{dmitriev} D. V. Dmitriev and V. Ya. Krivnov, {Phys. Rev. B} {\bf 73}, 024402
(2006).
\bibitem{kolezhuk} F. Heidrich-Meisner, I. P. McCulloch, and A. K. Kolezhuk,
Phys. Rev. B {\bf 80}, 144417 (2009).
\bibitem{nishimoto} S. Nishimoto, S.-L. Drechsler, R. Kuzian, J. Richter, and J. van den
Brink,
arXiv:1005.5500 (2010).
\bibitem{LW79} C.~P.~Landee and R.~D.~Willett, 
Phys.~Rev.~Lett. \textbf{43}, 463 (1979).
\bibitem{DRS82} C.~Dupas, J.~P.~Renard, J.~Seiden, and A.~Cheikh-Rouhou,  
Phys.~Rev.~B \textbf{25}, 3261 (1982).
\bibitem{TTN91} M.~Takahashi, P.~Turek, Y.~Nakazawa, M.~Tamura, 
K.~Nozawa, D.~Shiomi, M.~Ishikawa, and M.~Kinoshita, 
Phys.~Rev.~Lett.~\textbf{67}, 746 (1991);
Y.~Nakazawa, M.~Tamura, N.~Shirakawa, 
D.~Shiomi, M.~Takahashi, M.~Kinoshita, 
and M.~Ishikawa, Phys.~Rev.~B \textbf{46}, 8906 (1992).
\bibitem{TKI92} M.~Takahashi, M.~Kinoshita, and M.~Ishikawa, 
J.~Phys.~Soc.~Jpn.~\textbf{61}, 3745 (1992). 
\bibitem{SGM06} K.~Shimizu, T.~Gotohda, T.~Matsushita, N.~Wada, 
W.~Fujita, K.~Awaga, Y.~Saiga, and D.~S.~Hirashima, 
Phys.~Rev.~B \textbf{74}, 172413 (2006).
\bibitem{SLW79} D.~D.~Swank, C.~P.~Landee, and R.~D.~Willett, 
Phys.~Rev.~B \textbf{20}, 2154 (1979).
\bibitem{schmitt} M. Schmitt, J. M\'alek, S.-L. Drechsler, and H. Rosner
Phys. Rev. B {\bf 80}, 205111 (2009). 

\bibitem{haertel10} 
M. H\"{a}rtel, J. Richter, D. Ihle, and  S.-L. Drechsler,
      Phys. Rev. B {\bf 81},  174421 (2010).

\bibitem{magfeld} I. Junger, D. Ihle, J. Richter, and A. Kl\"{u}mper, Phys. Rev. B \textbf{70}, 104419
(2004).
\bibitem{ihle_new} I. Juh\'{a}sz Junger, D. Ihle, L. Bogacz, and W. Janke, Phys. Rev. B \textbf{77}, 174411
(2008).



\bibitem{kondo} J. Kondo and K. Yamaji, Prog. Theor. Phys. \textbf{47}, 807
(1972); H. Shimahara and S. Takada, J. Phys. Soc. Jpn. \textbf{60}, 2394
(1991); S. Winterfeldt and D. Ihle, Phys. Rev. B \textbf{56}, 5535 (1997).
\bibitem{rgm_new} W. Yu and S. Feng, Eur. Phys. J. B {\bf{13}}, 265 (2000); 
L. Siurakshina, D. Ihle, and R. Hayn, Phys. Rev. B \textbf{64}, 104406
(2001);
B.H. Bernhard, B. Canals, and C. Lacroix, Phys. Rev. B {\bf{66}},
024422 (2002); D. Schmalfu{\ss}, J. Richter, and D. Ihle, Phys. Rev. B \textbf{70}, 184412 (2004);
       I. Junger, D. Ihle, and J. Richter,
       Phys. Rev. B {\bf 72}, 064454 (2005);
       D. Schmalfu{\ss}, J. Richter, and D. Ihle, Phys. Rev. B {\bf 72}, 224405 (2005);
       D. Schmalfu{\ss}, R. Darradi, J. Richter, J.~Schulenburg,  and D.~Ihle, 
       Phys. Rev. Lett. {\bf 97}, 157201 (2006);
       I. Juh\'asz Junger, D. Ihle, and  J. Richter,
       Phys. Rev. B {\bf 80}, 064425 (2009).
\bibitem{antsyg} 
T.N.~Antsygina, M.I.~Poltavskaya, I.I.~Poltavsky, and K.A.~Chishko, Phys. Rev. B {\bf 77},  024407
(2008).
\bibitem{elk} W. Gasser, E. Heiner, and K.~Elk,  Greensche Funktionen in
Festk\"{o}rper- und Vielteilchenphysik, WILEY-VCH Verlag, Berlin (2001).
\bibitem{NR2} W.H.~Press, S.A.~Teukolsky, W.T.~Vetterling, and B.P.~Flannery, 
{\it Numerical Recipes in C, The Art of Scientific Computing} (Cambridge University Press, Cambridge,
1992).

\bibitem{yamada} M. Yamada  and M. Takahashi,
J. Phys. Soc. Jpn. \textbf{55}, 2024 (1986).


\bibitem{MRG00} V.~Markovich, E.~Rozenberg, G.~Gorodetsky, B.~Revzin,
J.~Pelleg, and I.~Felner, Phys.~Rev.~B \textbf{62}, 14186 (2000).
\bibitem{sznaid} J.~Sznajd,  Phys. Rev. B \textbf{64}, 052401 (2001).

\end{thebibliography}
\end{document}